\documentclass[11pt]{article}
\usepackage{moriond,epsfig, amssymb}

\def\be{\begin{equation}}
\def\ee{\end{equation}}
\def\bea{\begin{eqnarray}}
\def\eea{\end{eqnarray}}

\begin{document}
\vspace*{4cm}
\title{SEARCHES FOR NEW PHYSICS IN LEPTON FINAL STATES}

\author{ CATALIN I. CIOBANU\\
FOR THE CDF AND D\O\/ COLLABORATIONS }

\address{Department of Physics, University of Illinois at Urbana-Champaign,\\
1110 W. Green St., Urbana, IL 61801-3080, USA}

\maketitle\abstracts{
Final states containing charged leptons could provide some of the most 
distinctive signatures for observing physics beyond the Standard Model.
We present searches for new physics using $0.32-0.45$ fb$^{-1}$ of data
accumulated with the CDF II and D\O\/ detectors at the Tevatron. No significant 
evidence of a signal is found, and in most cases the tightest constraints to date 
are set on the exotic processes investigated.}

\section{Introduction}

The Standard Model (SM) of particle physics has withstood all experimental tests to date.
Despite its numerous successes, it is believed that physics beyond the Standard Model
should manifest itself at scales of the order of 1 TeV or higher. Some examples of theories
proposed to describe the new physics are Supersymmetry (SUSY), extra dimensions, 
or lepton/quark compositeness theories. We present the results obtained from 
testing the existence of several of these exotic models using the Tevatron Run II data. 
We focus only on the final states containing one or more charged leptons. 
Requiring the presence of leptons significantly reduces the jet backgrounds 
abundant at the Tevatron. In addition, both collider detectors have very good electron and muon 
triggering and identification capabilities. Tau lepton identification suffers from 
the large associated background (QCD-produced jets faking tau signal) but can in many 
cases contribute significantly to improving signal the acceptance.

\section{SUSY Searches}

The SUSY extensions of the SM assume a new symmetry such that for each SM
fermion there exists a SUSY boson, while for each SM boson there 
exists a SUSY fermion. Martin has written a comprehensive introduction to SUSY \cite{martin}.

\subsection{Chargino-Neutralino Searches}

\begin{figure}
{\begin{center}
\epsfig{figure=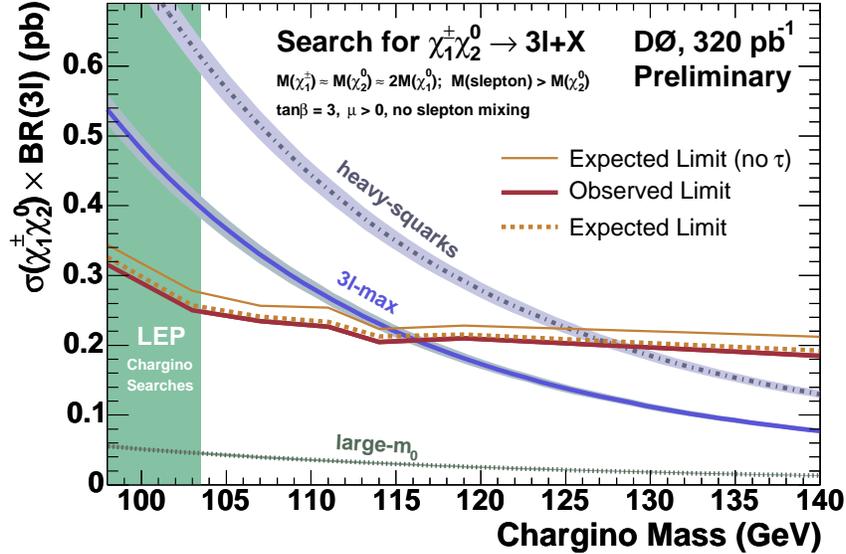,height=3in}
\end{center}}
\caption{D\O~limits on the total cross section for associated chargino and neutralino production with leptonic
final states. Also shown are the expectations from three particular SUSY scenarios: ``heavy-squarks'', 
``3$\ell$-max'', and ``large-$m_{0}$'', respectively (defined in the text).
\label{fig:d0cn}}
\end{figure}

In the simplest supersymmetric extension of the Standard Model, MSSM, the SUSY partners of the Higgs fields and 
SM $\gamma$, $W^{+}$, $W^{-}$, and $Z^0$ bosons mix to form two chargino ($\tilde{\chi}^{\pm}$) and four
neutralino $\tilde{\chi}^{0}$ mass eigenstates. The D\O~Collaboration has searched  \cite{d0cn} for the 
chargino-neutralino associated production $p\bar{p}\rightarrow \tilde{\chi}_{1}^{\pm} \tilde{\chi}_{2}^{0}$, 
where the lightest chargino $\tilde{\chi}_{1}^{\pm}$ and the
second-lightest neutralino $\tilde{\chi}_{2}^{0}$ decay as: 
$\tilde{\chi}_{1}^{\pm}\rightarrow \ell^{\pm}\nu\tilde{\chi}_{1}^{0}$ and
$\tilde{\chi}_{2}^{0}\rightarrow \ell^{\pm}\ell^{\mp}\tilde{\chi}_{1}^{0}$. 
The $\tilde{\chi}_{1}^{0}$ particle is expected to be stable if $R$-parity conservation 
is assumed \footnote{$R_{P}=(-1)^{3(B-L)+2s}$ where $B$, $L$, and $s$ denote the barionic and the leptonic
numbers, and the spin, respectively.}. The final state for this process consists of three charged 
leptons and missing energy from the neutrino and the stable $\tilde{\chi}_{1}^{0}$ particles. 
The dominant background sources which contribute to this final state are Drell-Yan dilepton 
production with an additional lepton coming from jet misidentification or conversion processes, or 
diboson $WZ$ production. The observed event yield agrees well with the expectation from SM sources 
alone, and upper limits are set on the chargino and neutralino cross section times branching 
fraction $\sigma \times$ BR $(3\ell)$. Figure \ref{fig:d0cn} shows this limit as function of chargino
mass, assuming no slepton mixing and the mSUGRA-inspired 
$m_{\tilde{\chi}_{1}^{\pm}}\approx m_{\tilde{\chi}_{2}^{0}}\approx 2m_{\tilde{\chi}_{1}^{0}}$. The three 
SUSY scenarios depicted in Fig. \ref{fig:d0cn} are: models with heavy squark masses and 
low slepton masses ($heavy$-$squark$ scenario), models with low sleptons masses in mSUGRA where
$m_{\tilde {\ell}}\gtrsim m_{\tilde{\chi}_{2}^{0}}$ (3$\ell$-$max$) , and models with 
large $m_{0}$ with the chargino and neutralino decaying via virtual gauge bosons ($large$ $m_{0}$, 
essentially unconstrained by the data). Adding $\tau$ leptons to this analysis was found to improve
these limits by about 2-3 GeV, depending on the SUSY scenario investigated \cite{d0cnt}.

\subsection{Squark and Sneutrino Searches}

\subsubsection{Scalar Top Searches in MSSM}
\begin{figure}
{\begin{center}
\epsfig{figure=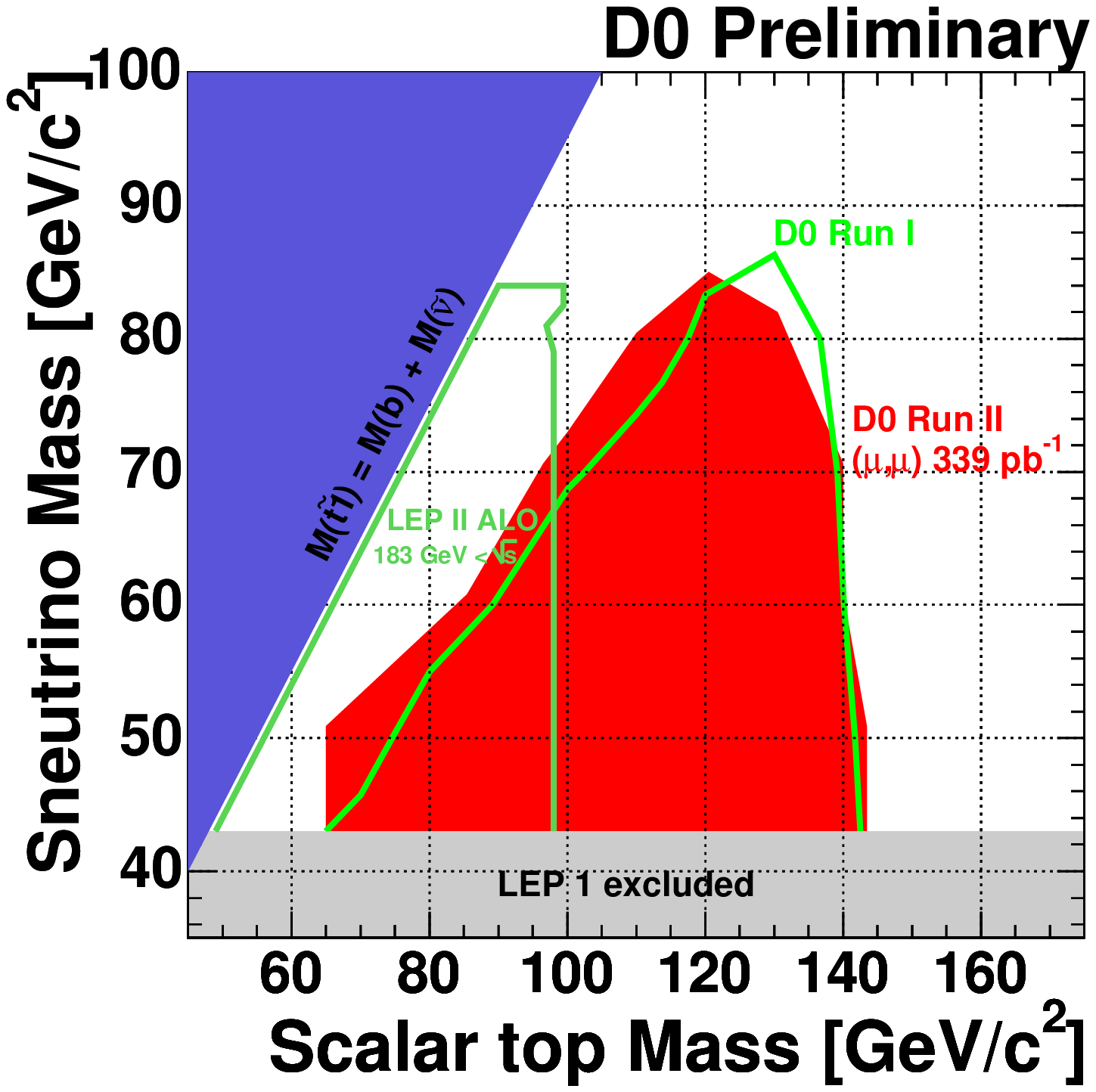,height=2.6in}
\epsfig{figure=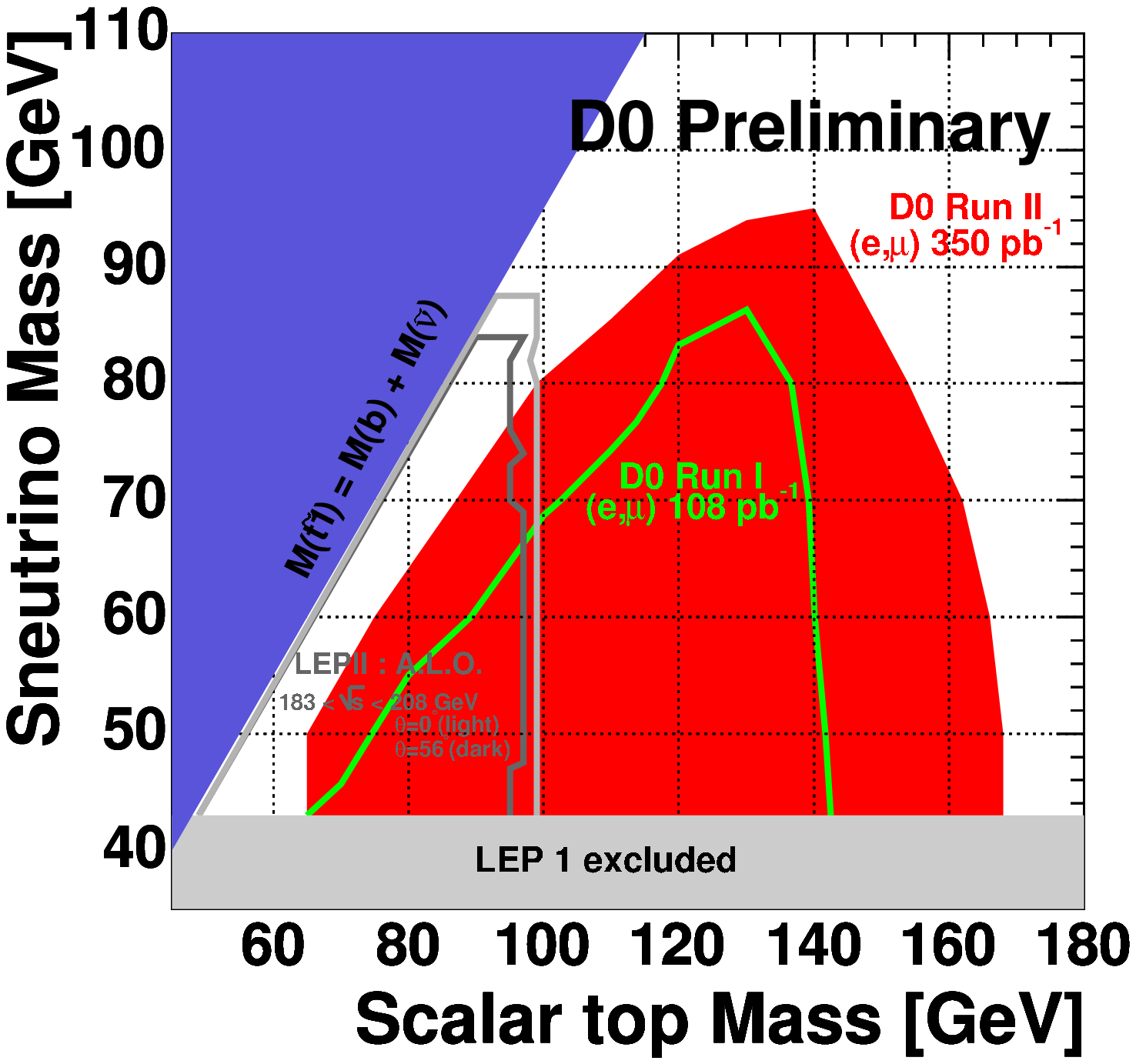,height=2.5in}
\end{center}}
\caption{D\O~ 95\% C.L. excluded region, compared to the LEP and Run I D\O~ results. 
\label{fig:d0stop}}
\end{figure}

There are two stop particles $\tilde{t}_{1}$ and $\tilde{t}_{2}$ corresponding to the
SM top quark. Due to the large mass of the latter, 
the effect of mixing on the $\tilde{t}$ mass is the largest among the squarks, 
and the lighter $\tilde{t}_{1}$ could be the lightest
of all squarks. The D\O~Collaboration has searched \cite{d0stop} for the stop pair production 
$p\bar{p}\rightarrow \tilde{t}_{1} \bar{\tilde{t}_{1}}\rightarrow 
b\tilde{\chi}^{+}\bar{b}\tilde{\chi}^{-}$. Only the decays 
$\tilde{\chi}^{\pm} \rightarrow e^{\pm}\tilde{\nu}, \mu^{\pm}\tilde{\nu}$ are considered, followed by the
sneutrino decays: $\tilde{\nu} \rightarrow \nu \tilde{\chi}_{1}^{0}$ where the $\tilde{\chi}_{1}^{0}$
is assumed to be the lightest supersymmetric particle (LSP). The final state is therefore
$b\bar{b}\ell^{+} \bar{\ell^{-}} \tilde{\nu} \bar{\tilde{\nu}}$, where the $\ell^{+} \bar{\ell^{-}}$ pairs studied are
$e^{\pm}\mu^{\mp}$ and $\mu^{\pm}\mu^{\mp}$. In addition to requiring a lepton pair, it is also required the
events contain at least one $b$-jet as identified from using silicon vertex detector information.
The dominant backgrounds to this analysis are top pair production, Drell-Yan ditau or dimuon production (for the 
$\mu^{\pm}\mu^{\mp}$ channel), jets misidentified as leptons ($e^{\pm}\mu^{\mp}$ channel), and diboson 
$WW$, $WZ$, and $ZZ$ processes. The observed event count agrees well with that expected from 
SM background sources alone. Exclusion regions are determined in the ($m_{\tilde{t}_{1}}, m_{\tilde{\nu}}$) plane 
(Fig. \ref{fig:d0stop}), and found to extend the previous tightest constraints obtained in searches at the LEP I,
LEP II and Run I Tevatron colliders.

\subsubsection{Scalar Top Searches in $R$-parity violating SUSY}
\begin{figure}
{\begin{center}
\epsfig{figure=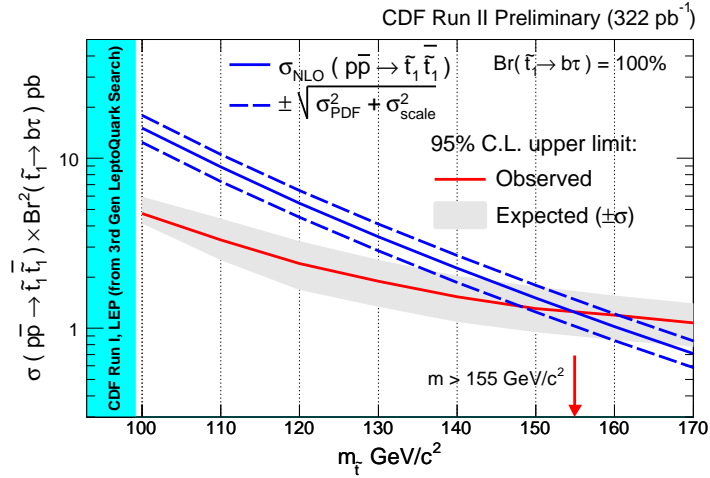,height=2.5in}
\end{center}}
\caption{CDF 95\% C.L. limit curve for the $\tilde{t}_{1} \bar{\tilde{t}_{1}}$ production, 
compared with NLO calculations (solid line). The dashed lines represent the uncertainties in the
theoretical calculation due to the choice of parton distribution functions and renormalization scale. 
\label{fig:cdfstop}}
\end{figure}

This search performed by the CDF Collaboration considers the stop pair production 
via $R$-parity conserving processes, followed by the $R$-parity violating decay of the two stop 
particles $\tilde{t}_{1} \bar{\tilde{t}_{1}}\rightarrow \bar{\tau} b \tau \bar{b}$.
The final state contains either an electron or a muon from the tau lepton decay
$\tau \rightarrow e\bar{\nu_{e}}\nu_\tau$ or $\mu\bar{\nu_{\mu}}\nu_\tau$, a hadronically decaying 
tau ($\tau_{h}$) lepton, and two or more jets. For this final state, the expected SM backgrounds are
events with a true $\ell \tau_{h}$ pair from Drell Yan $Z/\gamma$*$\rightarrow \tau\bar{\tau}+$jets,
$t\bar{t}$ and diboson, and events with fake $\ell \tau_{h}$ combinations from $W+$jets and QCD events.
As in the MSSM stop search, the event count is in good agreement with the background estimations.
The cross section times branching fractions upper limit dependence on the stop mass is shown in 
Fig. \ref{fig:cdfstop}. For the particular model considered, stop masses less than 155 GeV are 
excluded by the data at 95\% C.L. 

As an aside, we note that stop pair production is very similar to the
pair production of the third generation leptoquark (LQ$_{3}$), and therefore the limit curves 
shown in Fig.\ref{fig:cdfstop}
can be alternatively interpreted in terms of LQ$_{3}$ mass.

\subsubsection{Scalar Neutrino Searches in $R$-parity violating SUSY}

\begin{figure}
\begin{center}
\epsfig{figure=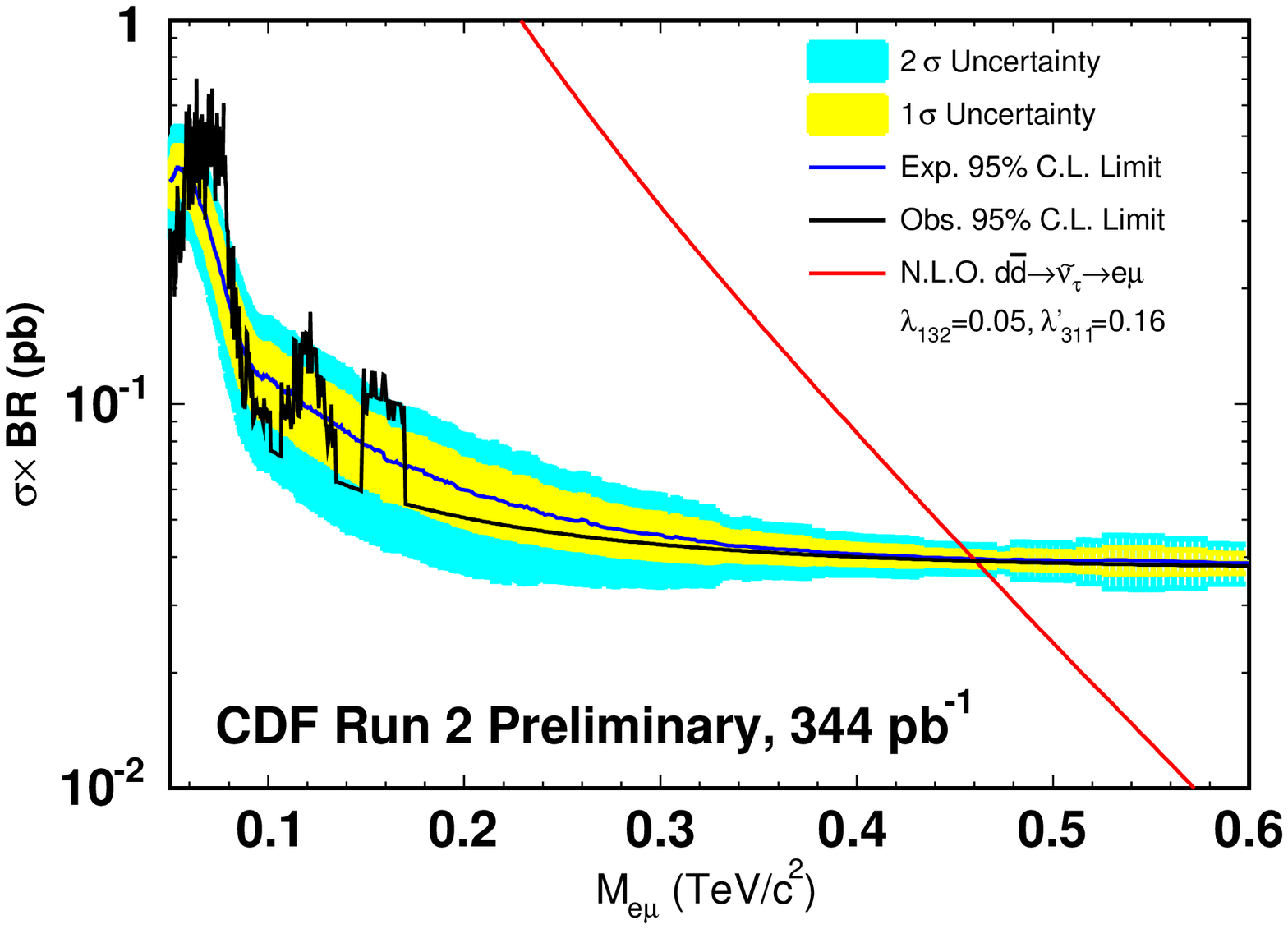,height=2.4in}
\epsfig{figure=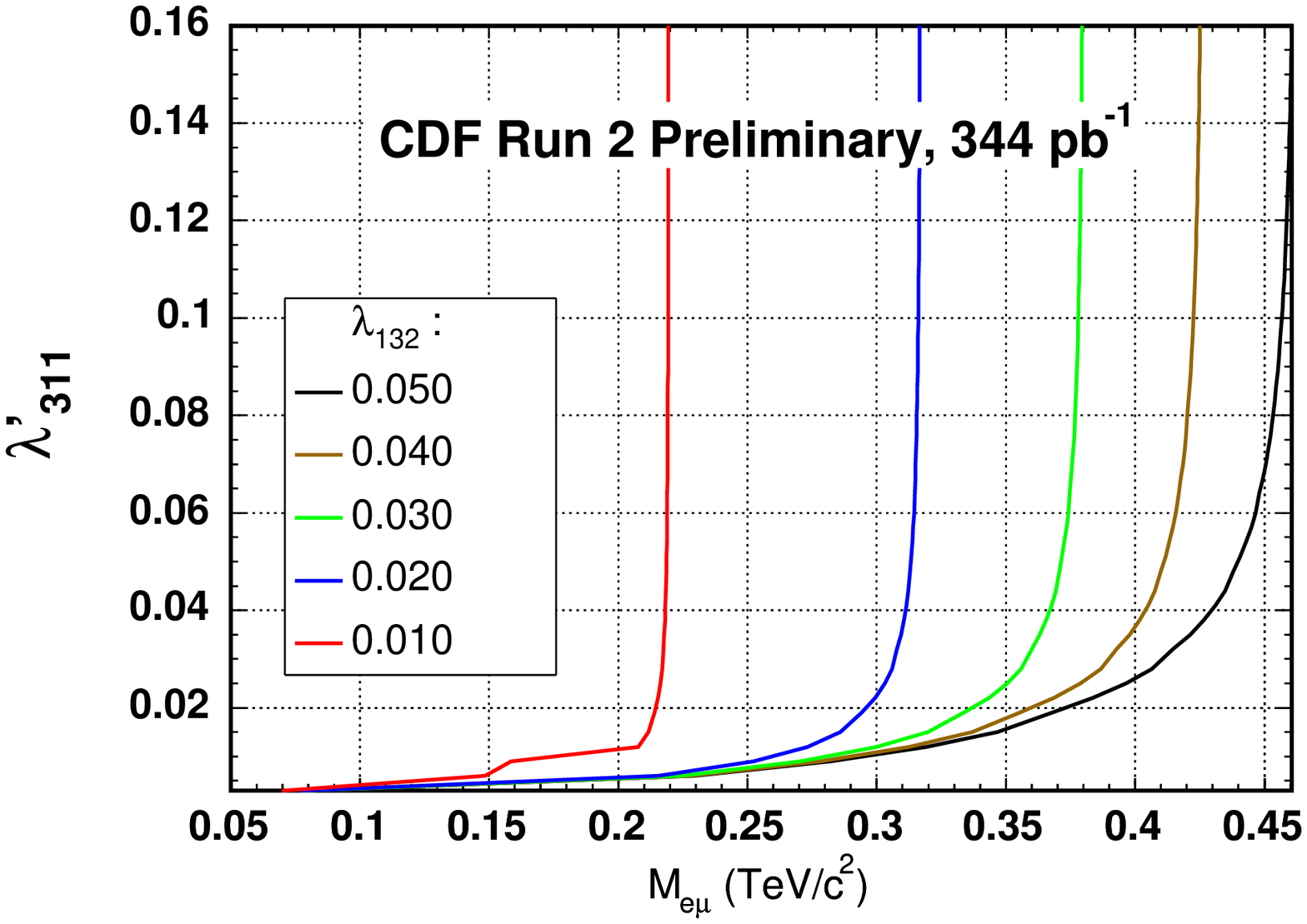,height=2.4in}
\end{center}
\caption{Left: CDF 95\% C.L. expected and observed upper limit curves for the sneutrino production and decay 
$d\bar{d}\rightarrow \tilde{\nu}_{\tau}\rightarrow e\mu$, compared to the NLO theoretical 
calculations (red line). Right: $\lambda^{\prime}_{311}-M_{e\mu}$ exclusion regions, for several 
$\lambda_{132}$ values.
\label{fig:cdfsnu}}
\end{figure}

The CDF Collaboration searched for the $R$-parity violating production and decay of the 
tau sneutrino: $d\bar{d}\rightarrow \tilde{\nu}_{\tau}\rightarrow e\mu$ \cite{kahn}. 
In the high reconstructed $e\mu$ mass region, the main backgrounds are top pair 
production $t\bar{t}$ and $WW$ diboson events. Very good agreement between the CDF data and 
the SM predictions is observed, with a p-value of 23\%.  
The theoretical cross section for the 
$d\bar{d}\rightarrow \tilde{\nu}_{\tau}\rightarrow e\mu$ depends on the strength of two
couplings in the SUSY Lagrangian, $\lambda_{132}$ and $\lambda^{\prime}_{311}$.
The left plot of Fig. \ref{fig:cdfsnu} shows the observed and expected upper limits and the theoretical 
cross section for this process calculated using the previous limits $\lambda_{132}=0.05$ and 
$\lambda^{\prime}_{311}=0.17$. The right plot in Fig. \ref{fig:cdfsnu}
shows the exclusion regions in the ($M_{e\mu}$, $\lambda^{\prime}_{311}$) as functions of the
assumed $\lambda_{132}$ value.
Compared to the previous results \cite{pdg1} there is a significant 
improvement in the limit obtained at the particular  ($\lambda$, $\lambda^{\prime}$) point, 
and more generally, $\tilde{\nu}_{\tau}$ masses are now excluded as a function of the two couplings. 

\section{$Z^{\prime}$ Boson, Contact Interaction and Excited Lepton Searches}

\subsection{Heavy $Z^{\prime}$ Searches}
\begin{figure}
\begin{center}
\epsfig{figure=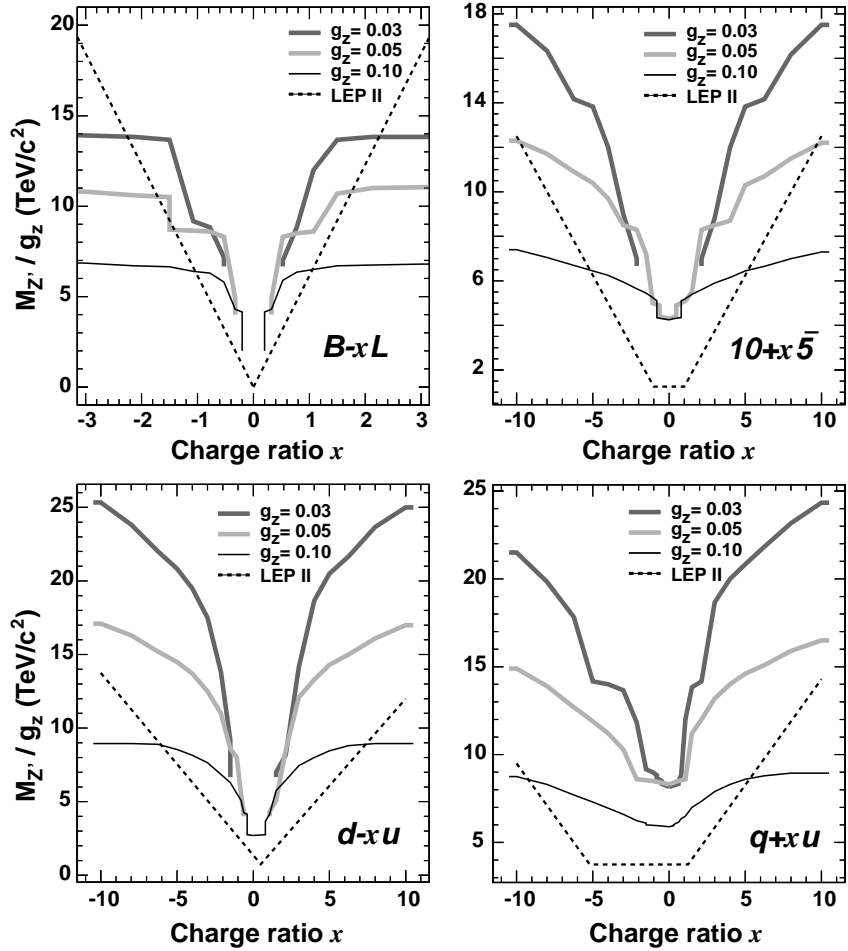,height=5in}
\end{center}
\caption{Exclusion contours for the $B-xL$, $10+x\bar{5}$, 
$d-xu$, and $q+xu$ $Z^{\prime}$ families. The dotted lines represent the
exclusion boundaries derived in Carena $et$ $al.$~$^{10}$ from the LEP II results $^{11}$. 
The region below each curve is excluded by our data at 95\% C.L.
Only models with $M_{Z^{\prime}}>$200 GeV are tested, which explains the gap at small $|x|$ for some models.
\label{fig:cdfzp}}
\end{figure}

Many extensions of the SM gauge group  predict the existence of electrically-neutral, 
massive gauge bosons commonly referred to as $Z^{\prime}$. The CDF Collaboration recently searched \cite{zee} for 
$Z^{\prime}\rightarrow e^{+}e^{-}$ decays in 0.45 fb$^{-1}$ of the data using the two dimensional
($M_{ee}$, $\cos\theta$*) distribution \footnote{$\theta$* is the scattering angle of the electron pair 
in the Collins-Soper frame.}. In the search region $M_{ee}\geq 200$ GeV, the main background to
$Z^{\prime}/Z/\gamma$* process are the events with jets misidentified as electrons, and to a smaller
extent diboson events. Good agreement is found between the data events and the expectation from 
SM sources. 95\% C.L. upper limits are set on the $Z^{\prime}$ mass for traditional models such as 
the sequential $Z^{\prime}$ (850 GeV), the canonical  superstring-inspired E$_{6}$ models: 
$Z_{\psi}$ (740 GeV) , $Z_{\chi}$ (725 GeV), $Z_{\eta}$ (745 GeV), $Z_{I}$ (650 GeV), $Z_{N}$ (710 GeV),
and $Z_{{\textrm sec}}$ (680 GeV) \footnote{The couplings for these models are summarized in Ciobanu $et$ $al.$ \cite{catutza}}. 
A more general parametrization for  $Z^{\prime}$ models was proposed \cite{CDDT}, where a
given $Z^{\prime}$ is specified by mass $M_{Z^{\prime}}$, gauge coupling $g_{z}$, and a 
certain charge ratio $x$. Figure \ref{fig:cdfzp} shows the 95\% C.L. exclusion
regions for the four families of models advocated in Carena $et$ $al.$ \cite{CDDT}, which, 
in the small $|x|$ and 
small $g_{z}$ regions, improve the $Z^{\prime}$ bounds derived from LEP II  contact interaction searches. 

\subsection{Contact Interaction Searches}
The D\O~Collaboration has used a similar technique to search \cite{d0coin} for evidence of four-fermion 
($q\bar{q}\mu\bar{\mu}$) contact interactions (CI) in 0.4 fb$^{-1}$ of data. An effective 
Lagrangian for the $q\bar{q}\mu\bar{\mu}$ CI can be written as:
$\sum_{q} \sum_{i,j=L,R}
 4\pi\eta
 \bar{\mu_i}\gamma^{\mu}\mu_i \bar{q_j}\gamma_{\mu}q_j/{\Lambda^2_{ij}}$,
where $\Lambda$ is the scale of the interaction, and $\eta = \pm\ 1$
determines the interference structure with the $Z/\gamma^{*}$ amplitudes~\cite{ci}.
A generation-universal interaction is assumed and lower
limits are measured for $\Lambda$ in eight helicity structure scenarios, shown in 
Table \ref{d0ci} .

\begin{figure}
\begin{center}
\epsfig{figure=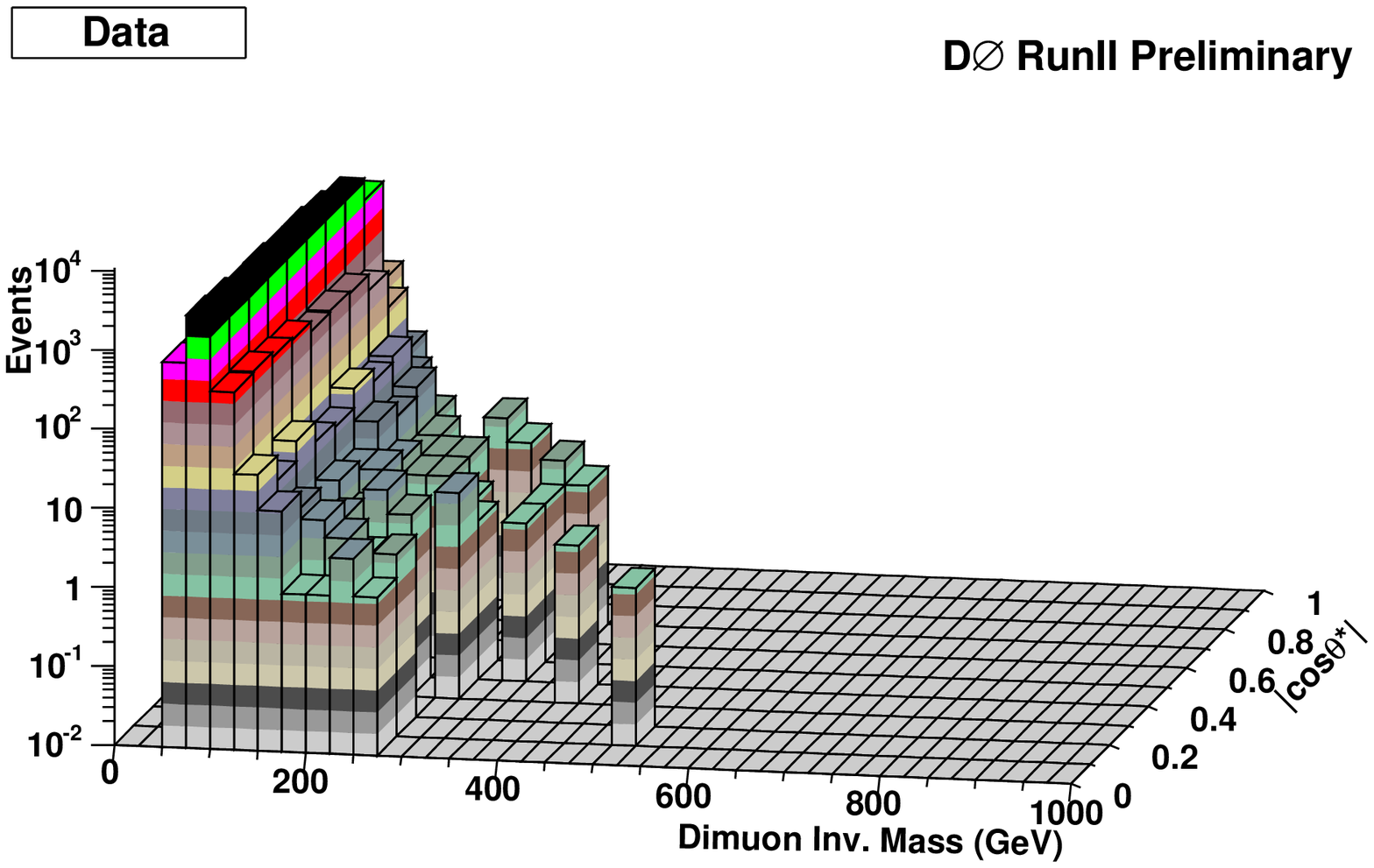,height=2.5in}
\epsfig{figure=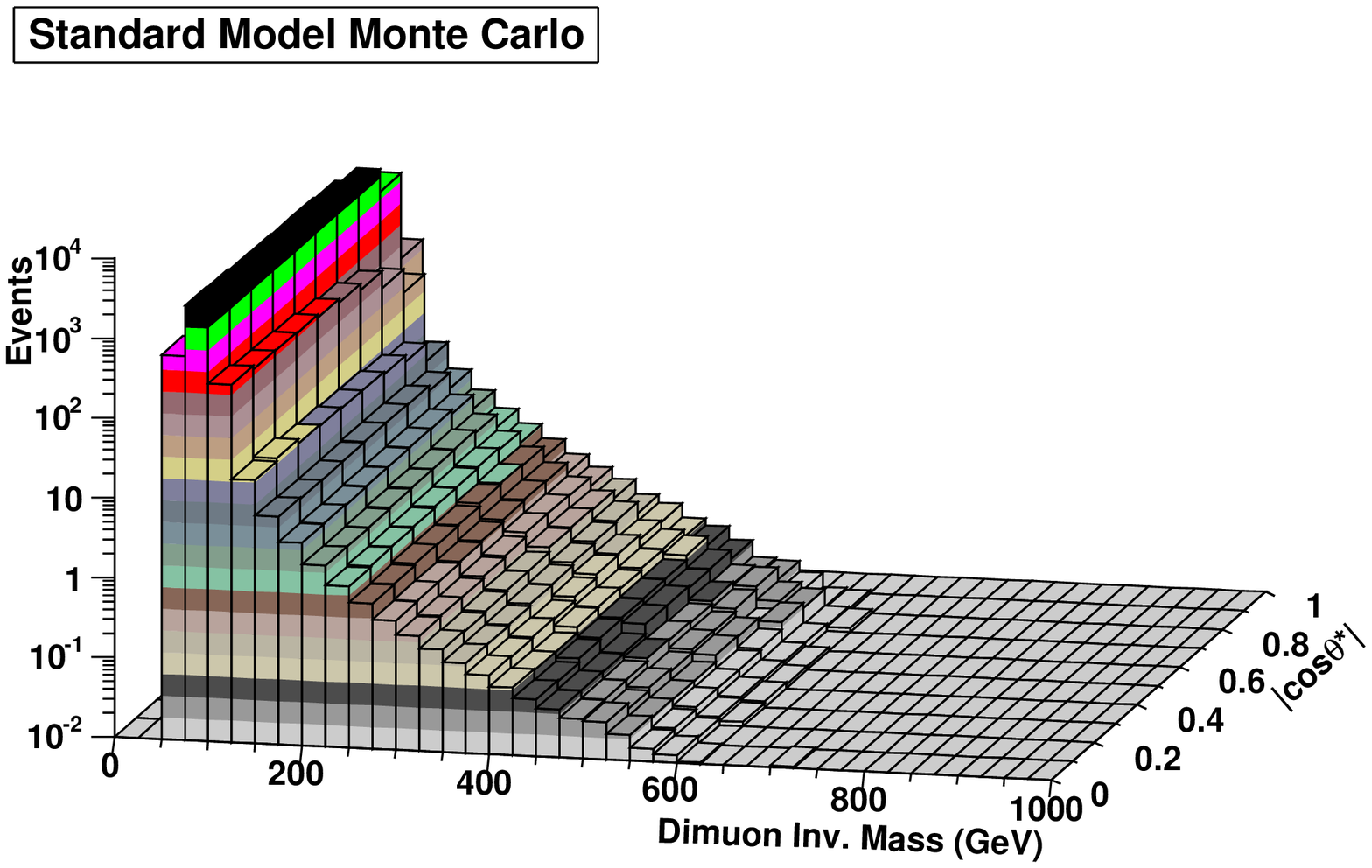,height=2.5in}
\end{center}
\caption{D\O: The bidimensional distribution ($M_{\mu\mu}$, $| \cos\theta$*$|$) for data events (upper plot) and SM 
Monte Carlo processes normalized to the effective luminosity (lower plot).
\label{fig:d0ci}}
\end{figure}

\begin{table}[tb] 
\centering\small 
\begin{tabular}{|c|c|c|c|c|c|c|c|c|c|c|}
\hline\hline
 Interaction & LL & RR & LR & RL & LL+RR & RL+LR & LL-RL & LR-RR & VV & AA \\
\hline
 $\Lambda^{+}_{q\mu}$ limit (TeV) & 4.19 & 4.15 & 5.32 & 5.31 & 5.05 & 6.45 & 4.87 & 5.07 &
 6.88 & 5.48 \\
\hline
 $\Lambda^{-}_{q\mu}$ limit (TeV) & 6.98 & 6.74 & 5.10 & 5.17 & 9.05 & 6.12 & 7.74 & 7.41 &
 9.81 & 9.76\\
\hline\hline
\end{tabular} 
\caption{D\O~95\% C.L. lower limits for the $q\bar{q}\mu\bar{\mu}$ contact 
interaction mass scales.}
\label{d0ci} 
\end{table} 

\subsection{Excited Lepton Searches}

\begin{figure}
\begin{center}
\epsfig{figure=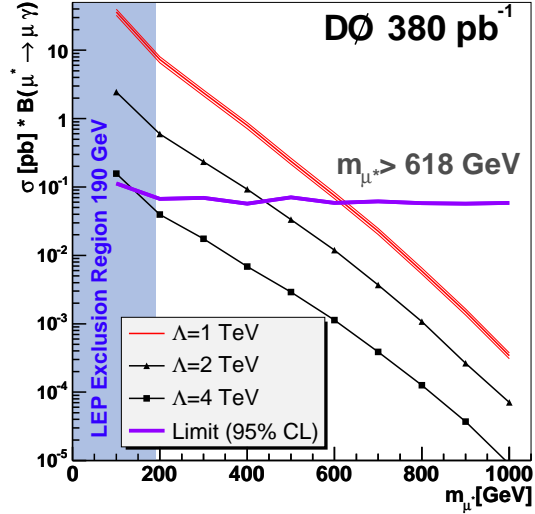,height=3in}
\end{center}
\caption{D\O: The 95\% C.L. upper limit on the product of the cross section and 
branching fraction, compared to the CI model predictions for different scales $\Lambda$.
For the case  $\Lambda=1$ TeV, the theoretical uncertainty of the model prediction is
also shown.
\label{fig:d0em}}
\end{figure}

In the lepton and quark compositeness model, a quark or a lepton is a bound state 
of three fermions, or of a fermion and a boson, which may have a spectrum of excited 
bound states. 
Recently, the D\O~Collaboration has 
searched for excited muons in 0.38 fb$^{-1}$ of data \cite{d0:em}. This analysis 
considers the single production of an excited muon $\mu$* in conjunction with a 
muon via four-point CI $q\bar{q}\mu\bar{\mu}$*, with the subsequent electroweak 
decay $\mu$*$\rightarrow \mu \gamma$. This decay mode competes with the 
decay through the CI mechanism, and its branching ratio depends on the ratio
$m_{\mu{\textrm *}}/\Lambda$ ($\Lambda$ denotes the compositeness scale). 
The main background contributing to the $\mu\mu\gamma$ final state are 
$Z/\gamma$*$\rightarrow \mu^{+}\mu^{-}(\gamma)$ decays, where the photon originates 
from initial or final state radiation. 
There are no observed data events in the signal region, consistent to the SM expectation. 
Figure \ref{fig:d0em} shows the limits on the product of cross section and branching ratio 
as functions of the mass of the excited muon. For $\Lambda=1$ TeV, $\mu$* masses up to 618 
GeV are excluded, at 95\% C.L. 

A similar search has been performed by the CDF Collaboration \cite{hkg2} in 0.37 fb$^{-1}$ of data.
In the gauge-mediated decay model,  $\mu$* masses within 100$-$410 GeV range are excluded, for
$f/\Lambda=10^{-2}$ GeV$^{-1}$ \footnote{$f$ is a phenomenological constant \cite{uli}.}. 
In the CI model \cite{hkg} $\mu$* masses in the 107$-$853 GeV interval are excluded,
for $m_{\mu{\textrm *}}=\Lambda$.
For the same $\mu$* model as considered in the D\O~search, the CDF $\mu$* mass limit is 696 GeV.

\section{Conclusions}
In conclusion, we present several topics from exotic searches in lepton final states
at the Tevatron. No significant excess for any of the signal models investigated is observed, 
and limits are set as functions of the relevant parameters of each model. The exotic physics 
program at the Tevatron is very mature and diverse, and both the D\O~ and CDF Collaborations
are in a good position to observe new physics phenomena if these are indeed 
accessible at the Tevatron.

\section*{Acknowledgments}
We would like to thank our colleagues at D\O~and CDF for helping us prepare this 
document. We are very grateful to the conference organizers and the European Union
for providing us with partial financial support for the stay in La Thuile.


\section*{References}
\begin{thebibliography}{99}
\bibitem{martin} S.~P.~Martin, ``A Supersymmetry Primer'', hep-ph/9709356.

\bibitem{d0cn} V.~Abazov~$et$~$al.$ (D\O~Collaboration),  Phys. Rev. Lett. {\bf 95} 151805 (2005).

\bibitem{d0cnt} D\O~Collaboration, Public notes D\O~Note 4740$-$CONF, D\O~Note 4741$-$CONF, 
D\O~Note 4742$-$CONF  (unpublished).

\bibitem{d0stop} D\O~Collaboration, Public notes D\O~Note 4866$-$CONF and 
D\O~Note 5050$-$CONF  (unpublished).

\bibitem{cdfstop}CDF~Collaboration, Public note CDF 7835 (unpublished).

\bibitem{kahn} A.~Abulencia~$et$~$al.$ (CDF~Collaboration),  Phys. Rev. Lett. {\bf 96} 211802 (2006).

\bibitem{pdg1} S. Eidelman et al. (Particle Data Group), Phys. Lett. B 592, 1 (2004).

\bibitem{zee} A.~Abulencia~$et$~$al.$ (CDF~Collaboration),  Phys. Rev. Lett. {\bf 96} 211801 (2006).

\bibitem{catutza} C.~Ciobanu {\it et al.}, FERMILAB-FN-0773-E (2005).

\bibitem{CDDT} M.~Carena {\it et al.}, Phys. Rev. D {\bf 70}, 093009 (2004).

\bibitem{lep} D.~Abbaneo {\it et al.} (the LEP Collaborations) and N.~de~Groot {\it et al.} (the
SLD Collaboration), CERN-EP/2003-091 (2003).

\bibitem{d0coin} D\O~Collaboration, Public note D\O~Note 4922$-$CONF.

\bibitem{ci}E.~J.~Eichten, K.~D.~Lane, M.~E.~Peskin, Phys. Rev. Lett. {\bf 50}, 811 (1983).

\bibitem{d0:em} V.~M.~Abazov~$et$~$al.$ (D\O~Collaboration), Phys. Rev. D {\bf 73}, 111102(R) (2006). 

\bibitem{hkg2}CDF~Collaboration, Public note CDF 8145 (unpublished).

\bibitem{uli}U.~Baur, M.~Spira and P.~M.~Zerwas, Phys. Rev. D {\bf 42}, 815 (1990), 
and references therein; E. Boos et al., Phys. Rev. D {\bf 66}, 013011 (2002), 
and references therein. 

\bibitem{hkg}D.~Acosta~$et$~$al.$ (CDF~Collaboration), Phys. Rev. Lett  {\bf 94}, 101802 (2005).

\end{thebibliography}
\end{document}